\documentclass[pra,aps,twocolumn,epsfig,superscriptaddress,showpacs]{revtex4}
\usepackage[dvips]{graphicx}
\def\QED{\leavevmode\unskip\penalty9999 \hbox{}\nobreak\hfill
     \quad\hbox{\leavevmode  \hbox to.77778em{%
               \hfil\vrule   \vbox to.675em%
               {\hrule width.6em\vfil\hrule}\vrule\hfil}}
     \par\vskip3pt}
\def\qed{\leavevmode\unskip\penalty9999 \hbox{}\nobreak\hfill
     \quad\hbox{\leavevmode  \hbox to.77778em{%
               \hfil\vrule   \vbox to.675em%
               {\hrule width.6em\vfil\hrule}\vrule\hfil}}
     \par\vskip3pt}

\def\ibb #1{\leavevmode\hbox{\kern.3em\vrule
     height 1.5ex depth -.1ex width .4pt\kern-.3em\rm#1}}

\newcommand{\tr}{\mbox{Tr}}

\begin{document}

\title{Geometric interpretation for A-fidelity and its relation with Bures fidelity}

\author{Zhihao Ma}
\affiliation{Department of Mathematics, Shanghai Jiaotong
University, Shanghai, 200240, P.R.China}

\author{Fu-Lin Zhang}
\affiliation{Theoretical Physics Division, Chern Institute of
Mathematics, Nankai University, Tianjin, 300071, P.R.China}

\author{Jing-Ling Chen}
\email[Email:]{chenjl@nankai.edu.cn}\affiliation{Theoretical Physics
Division, Chern Institute of Mathematics, Nankai University,
Tianjin, 300071, P.R.China}

\date{\today}

\begin{abstract}
 A geometric interpretation for the A-fidelity between two states
of a qubit system is presented, which leads to an upper bound of the
Bures fidelity. The metrics defined based on the A-fidelity are
studied by numerical method. An alternative generalization of the
A-fidelity, which has the same geometric picture, to a $N$-state
quantum system is also discussed.
\end{abstract}

\pacs{03.67.-a, 03.65.Ta}


\maketitle



\section{Introduction}
The concept of fidelity plays an important role in quantum
computation and quantum information
\cite{Book,Vedral,Barnum,Bures,Uhlmann,Uhlmann1,Uhlmann2}. It is a
measurement of \textit{closeness} between two states. The most
well-known definition is the Bures fidelity between two states
$\rho_1$ and $\rho_2$, given by
\cite{Bures,Uhlmann,Uhlmann1,Uhlmann2},
\begin{eqnarray}
F_B(\rho_1,\rho_2)= \biggr[\tr(\sqrt{\sqrt{\rho_1} \rho_2
\sqrt{\rho_1}})\biggr]^2.
\end{eqnarray}
Assume that $\rho_1=| \phi \rangle\langle \phi |$ and $\rho_2= |
\varphi \rangle\langle \varphi |$ are two pure states, then
$F_B(\rho_1,\rho_2)= | \langle \phi | \varphi \rangle |^2$. Another
generalization of the usual transition probability from pure states
to mixed states is the quantum affinity \cite{Fa,Luo}
\begin{eqnarray}
F_A(\rho_1,\rho_2)= [\tr(\sqrt{\rho_1} \sqrt{\rho_2})]^2,
\end{eqnarray} which has many the same fundamental properties as
the Bures fidelity \cite{Luo}. For two pure state $\rho_1=| \phi
\rangle\langle \phi |$ and $\rho_2= | \varphi \rangle\langle \varphi
|$, we have $F_A(\rho_1,\rho_2)= | \langle \phi | \varphi \rangle
|^4$. Since it is denoted as $F_A$ in \cite{Fa} by Raggio, we would
like to call it A-fidelity in this report for its close relation and
comparability with the Bures fidelity.

It is always interesting and useful to find an intuitive geometric
picture of a concept in quantum information. For instance, the trace
distance between two single qubit states has a simple geometric
interpretation as half of the ordinary Euclidean distance between
points on the Bloch sphere \cite{Book}. Chen and his collaborators
\cite{ChenPRA1} provided a geometric picture for the Bures fidelity
for the qubit case. This brief report is aimed at putting forward a
geometric interpretation for the A-fidelity. At first, we would like
to review the result in the Bures case.

\begin{figure}
\includegraphics[width=7cm]{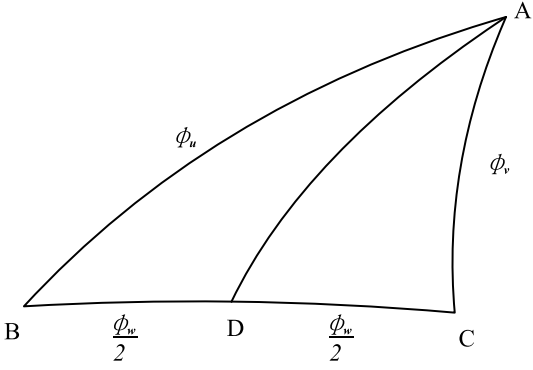} \\
 \caption{The
hyperbolic triangle $ABC$. Its three sides are
$|AB|=\phi_{\mathbf{u}}= \tanh ^{-1} |\mathbf{u}|$,
$|AC|=\phi_{\mathbf{v}}= \tanh ^{-1} |\mathbf{v}|$, and
$|BC|=\phi_{\mathbf{w}}= \tanh ^{-1} |\mathbf{w}|$. $D$ is the
midpoint of the side $BC$. The angle between $AB$ and $AC$ is equal
to $\pi - \cos^{-1} (\mathbf{\hat{u}} \cdot \mathbf{\hat{v}}) $}
\label{fig}
\end{figure}

The state of a qubit is described by a $2 \times 2$ density matrix
as
\begin{eqnarray} \label{state}
\rho(\mathbf{n})= \frac{1}{2}(\mathbf{1}+\vec{\sigma}\cdot
\mathbf{n}),   |\mathbf{n}|\leq 1
\end{eqnarray}
where $\mathbf{1}$ is the $2 \times 2$ unit matrix,
$\vec{\sigma}=(\sigma_1, \sigma_2, \sigma_3 )$ are the Pauli
matrices in vector notation, and $\mathbf{n}$ is the
three-dimensional Bloch vector. If $|\mathbf{n}|=1$, Eq.
(\ref{state}) corresponds to a pure state, otherwise a mixed state.
Chen $\textit{et al.}$ introduced the hyperbolic parameter
$\phi_{\mathbf{n}}$ , called \textit{rapidity}, to represent the
Bloch vector by the equation
\begin{eqnarray} \label{hpbl}
\mathbf{n}=\mathbf{\hat{n}} \tanh \phi_{\mathbf{n}},
\end{eqnarray}
where $\mathbf{\hat{n}}$ is the unit vector in the direction of
$\mathbf{n}$. The state in Eq. (\ref{state}) is represented as
\begin{eqnarray} \label{rep}
\rho(\mathbf{n})= \frac{L(\mathbf{n})}{2 \cosh \phi_{\mathbf{n}}},
\end{eqnarray}
where $L(\mathbf{n})$ is the Lorentz boost matrix,
\begin{eqnarray}
L(\mathbf{n})&=&  \exp (\frac{\varphi_{\mathbf{n}}}{2} \vec{\sigma}
\cdot \mathbf{\hat{n}}) \nonumber\\ &=& \mathbf{1} \cosh
\frac{\varphi_{\mathbf{n}}}{2}+\vec{\sigma} \cdot \mathbf{\hat{n}}
\sinh \frac{\varphi_{\mathbf{n}}}{2}, \\
\phi_{\mathbf{n}} &=& \varphi_{\mathbf{n}}/2. \nonumber
\end{eqnarray}
The Bloch vector $\mathbf{n}$ corresponds to a relativistically
admissible velocity, with the vacuum speed of light $c=1$. Then, the
Bures fidelity between two states
\begin{eqnarray}  \label{state2}
\rho_1= \frac{1}{2}(\mathbf{1}+\vec{\sigma}\cdot \mathbf{u}), \nonumber\\
\rho_2= \frac{1}{2}(\mathbf{1}+\vec{\sigma}\cdot \mathbf{v}),
\end{eqnarray}
is given by
\begin{eqnarray}\label{GeoFb}
F_B(\rho_1,\rho_2)=\frac{\cosh(\phi_{\mathbf{w}}/2)}{\cosh
\phi_{\mathbf{u}}} \frac{\cosh(\phi_{\mathbf{w}}/2)}{\cosh
\phi_{\mathbf{v}}}.
\end{eqnarray}
Here, $\mathbf{w}$ is the Einstein sum of the two relativistically
admissible velocities $\mathbf{u}$ and $\mathbf{v}$
\begin{eqnarray}\label{Vsum}
\mathbf{w}= \mathbf{u}\oplus \mathbf{v}= \frac{1}{1+\mathbf{u} \cdot
\mathbf{v}} \biggr[\mathbf{u}+ \frac{1}{\gamma_{\mathbf{u}}} \mathbf{v} +
\frac{\gamma_{\mathbf{u}}}{1+\gamma_{\mathbf{u}}} (\mathbf{u} \cdot
\mathbf{v}) \mathbf{u}\biggr],
\end{eqnarray}
where $\gamma_{\mathbf{u}}=1/\sqrt{1- |\mathbf{u}|^2} = \cosh
\phi_{\mathbf{u}}$ is the Lorentz factor. The rapidity
$\phi_{\mathbf{w}}$ satisfies the Cosine law of hyperbolic geometry
\begin{eqnarray} \label{cos}
\cosh \phi_{\mathbf{w}}=\cosh \phi_{\mathbf{u}} \cosh
\phi_{\mathbf{v}} (1+ \mathbf{\hat{u}} \cdot \mathbf{\hat{v}}  \tanh
\phi_{\mathbf{u}} \tanh \phi_{\mathbf{v}}).\ \
\end{eqnarray}
The hyperbolic angles $\{ \phi_{\mathbf{u}}, \phi_{\mathbf{v}}
,\phi_{\mathbf{w}} \}$ form a hyperbolic triangle, which is shown in
Fig.\ref{fig} by Chen \textit{et al.} \cite{ChenPRA1}.

\section{Geometric meaning of A-fidelity}

\emph{ Theorem 1.} The A-fidelity between two states $\rho_1$ and
$\rho_2$ in Eq. (\ref{state2}) equals to
\begin{eqnarray}\label{GeoFa}
F_A(\rho_1,\rho_2)&=&\frac{\cosh(\phi_{\mathbf{w}}/2)}{\cosh
\phi_{\mathbf{u}}} \frac{\cosh(\phi_{\mathbf{w}}/2)}{\cosh
\phi_{\mathbf{v}}} \cos^2 \frac{\delta}{2}, \nonumber \\
&=&F_B(\rho_1,\rho_2)\cos^2 \frac{\delta}{2},
\end{eqnarray}
where $\mathbf{w}$ is defined in Eq. (\ref{Vsum}), and $\delta$ is
the defect of the triangle $\triangle ABC$ in Fig.\ref{fig} which
satisfies \cite{Geo1,Geo2,Geo3}
\begin{eqnarray} \label{cosdel}
\cos \frac{\delta}{2}= \frac{1+\cosh \phi_{\mathbf{u}} +\cosh
\phi_{\mathbf{v}} +\cosh \phi_{\mathbf{w}} }{4\cosh
\frac{\phi_{\mathbf{u}}}{2}\cosh \frac{\phi_{\mathbf{v}}}{2}\cosh
\frac{\phi_{\mathbf{w}}}{2}}.
\end{eqnarray}

\emph{Proof.} In terms of the hyperbolic parameter, we can transform
Eq. (\ref{rep}) into
\begin{eqnarray}
\sqrt{\rho(\mathbf{n})}=\frac{\cosh(\phi_{\mathbf{n}}/2)}{\sqrt{2
\cosh \phi_{\mathbf{n}}}}[\mathbf{1} + \vec{\sigma} \cdot
\mathbf{\hat{n}} \tanh(\phi_{\mathbf{n}}/2)].
\end{eqnarray}
Then,
\begin{eqnarray}
\sqrt{\rho_1}\sqrt{\rho_2}=\frac{\cosh(\phi_{\mathbf{u}}/2)}{\sqrt{2
\cosh \phi_{\mathbf{u}}}}[\mathbf{1} + \vec{\sigma} \cdot
\mathbf{\hat{u}} \tanh(\phi_{\mathbf{u}}/2)] \nonumber \\
\times
\frac{\cosh(\phi_{\mathbf{v}}/2)}{\sqrt{2
\cosh \phi_{\mathbf{v}}}}[\mathbf{1} + \vec{\sigma} \cdot
\mathbf{\hat{v}} \tanh(\phi_{\mathbf{v}}/2)].
\end{eqnarray}
From the relations
\begin{eqnarray}
\tr(\mathbf{1}) &=& 2,  \nonumber \\
\tr(\vec{\sigma} \cdot \mathbf{\hat{n}}) &=& 0, \\
( \vec{\sigma} \cdot
\mathbf{\hat{u}})( \vec{\sigma} \cdot \mathbf{\hat{v}})
&=& \mathbf{\hat{u}} \cdot \mathbf{\hat{v}} \mathbf{1} + i \vec{\sigma} \cdot (\mathbf{\hat{u}} \times \mathbf{\hat{v}} ), \nonumber
\end{eqnarray}
we have
\begin{eqnarray}\label{trace}
&&\tr(\sqrt{\rho_1}\sqrt{\rho_2})=        \nonumber \\
&&  \ \ \ \ \ \frac{\cosh \frac{ \phi_{\mathbf{u}}}{2} \cosh\frac{ \phi_{\mathbf{v}}}{2}}{\sqrt{
\cosh \phi_{\mathbf{u}}  \cosh \phi_{\mathbf{v}}}} \biggr[1+ \mathbf{\hat{u}} \cdot \mathbf{\hat{v}} \tanh\frac{ \phi_{\mathbf{u}}}{2}\tanh\frac{ \phi_{\mathbf{v}}}{2}\biggr].  \ \ \ \ \
\end{eqnarray}
From the Cosine law Eq. (\ref{cos}), one can easily obtain
\begin{eqnarray}
\mathbf{\hat{u}} \cdot \mathbf{\hat{v}} =\frac{\cosh \phi_{\mathbf{w}} -\cosh \phi_{\mathbf{u}} \cosh \phi_{\mathbf{v}} }{\sinh \phi_{\mathbf{u}} \sinh \phi_{\mathbf{v}}}.
\end{eqnarray}
Substituting it into Eq. (\ref{trace}), along with some relations
between trigonometric functions of hyperbolic geometry, we derive
Eq. (\ref{GeoFa}), which ends the proof.

\section{Upper bound of Bures fidelity}

In the above section, we obtain a geometric picture of the
A-fidelity in hyperbolic geometry. It is also shown that, for a
qubit system, the factor connecting the A-fidelity and Bures
fidelity has a simple geometric meaning. Inserting Eq. (\ref{hpbl})
 into Eqs. (\ref{GeoFb}) and (\ref{cosdel}), one obtains
\begin{eqnarray}
F_B&=&\frac{1}{2}[1+\mathbf{u} \cdot \mathbf{v} + \sqrt{1-|\mathbf{u}|^2} \sqrt{1-|\mathbf{v}|^2}],\\
\cos^2 \frac{\delta}{2}&=&\frac{[2F_B+ \sqrt{1-|\mathbf{u}|^2} + \sqrt{1-|\mathbf{v}|^2} ]^2}{4(1+\sqrt{1-|\mathbf{u}|^2})(1+\sqrt{1-|\mathbf{v}|^2}) F_B}. \ \ \ \ \
\end{eqnarray}
Let
\begin{eqnarray}
f(x)&=&4[(1+\sin \alpha)(1+\sin \beta)-1] x^2 \nonumber \\
& & -4(\sin \alpha+ \sin \beta)x -(\sin \alpha+ \sin \beta)^2,
\end{eqnarray}
where $\alpha = \cos^{-1} |\mathbf{u}|$ and $\beta = \cos^{-1}
|\mathbf{v}|$. Then, we have
\begin{eqnarray}
F_B- \cos^2 \frac{\delta}{2} =\frac{1}{4(1+\sin \alpha)(1+\sin\beta)F_B} f(F_B).
\end{eqnarray}
Given a pair of $|\mathbf{u}|$ and $|\mathbf{v}|$, the corresponding
Bures fidelity satisfies
\begin{eqnarray}
0\leq \frac{1- \cos (\alpha + \beta)}{2} \leq F_B \leq \frac{1+ \cos (\alpha - \beta)}{2} \leq 1.
\end{eqnarray}
Considering
\begin{eqnarray}
f(0)=-(\sin \alpha + \sin \beta)^2 \leq 0, \nonumber \\
f(1)=-(\sin \alpha - \sin \beta)^2 \leq 0,  \\
4[(1+ \sin \alpha)(1+ \sin \beta)-1] \geq 0, \nonumber
\end{eqnarray}
we have
\begin{eqnarray} \label{boundFb}
F_B \leq \cos^2 \frac{\delta}{2}.
\end{eqnarray}
Only when $\mathbf{u}=\mathbf{v}$ or $|\mathbf{u}|=|\mathbf{v}|=1$,
one has $F_B = \cos^2 \frac{\delta}{2}$. This result shows that the
defect of $\triangle ABC$ in Fig. \ref{fig} associates with an upper
bound of the Bures fidelity.

On the other hand, from our result in Eq. (\ref{GeoFa}), one can
obtain
\begin{eqnarray}\label{ineq1}
F_A= F_B \cos^2\frac{\delta}{2} \leq F_B.
\end{eqnarray}
$F_A (\rho_1,\rho_2)= F_B (\rho_1,\rho_2)$, when $\rho_1$ commutes
with $\rho_2$. Then, $\mathbf{u} \texttt{//}\mathbf{v}$, $\angle BAC
= \pi$, the triangle $\triangle ABC$ becomes a line and $\delta=0$.

From Eqs. (\ref{ineq1}) and (\ref{boundFb}), we can immediately
obtain
\begin{eqnarray}
F_B^2 \leq F_A  \leq F_B,
\end{eqnarray}
which coincides with the result in \cite{Fa}.

The density matrix of a qunit ($N$-state system) \cite{qnit1,qnit2}
can be written as
\begin{eqnarray} \label{rhoN}
\rho(\mathbf{m})=\frac{1}{N} \biggr[\mathbf{1}_N+
\sqrt{\frac{N(N-1)}{2}} \vec{\lambda} \cdot \mathbf{m} \biggr],
\end{eqnarray}
where $\mathbf{1}_N$ denotes the $N \times N$ unit matrix,
$\vec{\lambda}= (\lambda_1, \lambda_2, ... ,\lambda_{N^2-1})$ are
the generators of $SU(N)$, and $\mathbf{m}$ is the
$(N^2-1)$-dimensional Bloch vector. It is expected that $\cos^2
\frac{\delta}{2}$ given in Eq. (\ref{cosdel}) is always the upper
bound of the Bures fidelity for arbitrary $N$. We have tested the
relation (\ref{boundFb}), for $N=3$ and 4, by numerical computation.
The results of $10^5$ random pairs of states show it is indeed the
case.
%

\section{Metrics related to A-fidelity}

Fidelity by itself is not a metric,
but there are many metrics built up from Bures fidelity to measure the distance between two quantum states.
The most famous ones are  known in the literature as the {\em Bures angle}, the {\em Bures metric}, and the
\emph{gold metric} \cite{Metric}, given by
\begin{eqnarray}
A(\rho,\sigma)&=&\arccos{\sqrt{F_B(\rho,\sigma)}},\nonumber\\
B(\rho,\sigma)&=&\sqrt{2-2\sqrt{F_B(\rho,\sigma)}},\\
C(\rho,\sigma)&=&\sqrt{1-F_B(\rho,\sigma)}. \nonumber
\end{eqnarray}
They satisfy the the following four axioms of a metric:

(M1). $d(x,y) \ge 0$ for all states $x$ and $y $;

(M2). $d(x,y)=0$ if and only if  $x=y$;

(M3). $d(x,y)=d(y,x)$ for all states $x$ and $y$;

(M4). The triangle inequality: $d(x,y)\leq d(x,z)+d(y,z)$ for all
states $x, y$ and $z$.

For a qubit system, it is easy to prove a common ground of the above
three metrics
\begin{eqnarray}
& &\lim_{\mathbf{u},\mathbf{v} \rightarrow 0} A(\rho_1,\rho_2)
=\lim_{\mathbf{u},\mathbf{v} \rightarrow 0} B(\rho_1,\rho_2)
=\lim_{\mathbf{u},\mathbf{v} \rightarrow 0} C(\rho_1,\rho_2) \ \  \nonumber \\
& & \ =\frac{1}{2}| \mathbf{u}-\mathbf{v} |.
\end{eqnarray}
Namely, when $|\mathbf{u}|$ and $|\mathbf{v}|$ approach $0$, all the
above limits are the trace distance of two states of a qubit
\cite{Book}. It is easy to prove
\begin{eqnarray}
\lim_{\mathbf{u},\mathbf{v} \rightarrow 0}F_A(\rho_1,\rho_2)
=\lim_{\mathbf{u},\mathbf{v} \rightarrow 0}F_B(\rho_1,\rho_2)
=1-\frac{1}{4}|\mathbf{u}-\mathbf{v}|^2.
\end{eqnarray}
This result suggests, if we introduce,
\begin{eqnarray}
\mathcal{A}(\rho,\sigma)&=&\arccos{\sqrt{F_A(\rho,\sigma)}},\nonumber\\
\mathcal{B}(\rho,\sigma)&=&\sqrt{2-2\sqrt{F_A(\rho,\sigma)}},\\
\mathcal{C}(\rho,\sigma)&=&\sqrt{1-F_A(\rho,\sigma)}, \nonumber
\end{eqnarray}
their limits are also the trace distance when the two states
approach the center of the Bloch sphere. Furthermore, they are three
well-defined metrics of qubit and qunit states. Actually Raggio
\cite{Fa} has analytically proved that $\mathcal{B}(\rho,\sigma)$ is
a metric. For $\mathcal{A}(\rho,\sigma)$ and
$\mathcal{C}(\rho,\sigma)$, we have numerically verified the
triangle inequality by using random $10^5$ sets of states for $2$,
$3$ and $4$-dimensional system separately. The analytic proof will
be given in the subsequent investigation.

\section{conclusion and discussion}
We have proposed a geometric observation for the A-fidelity between
two states of a qubit in terms the hyperbolic parameters introduced
in \cite{ChenPRA1}. The A-fidelity is shown as the product of Bures
fidelity and $\cos^2 \frac{\delta}{2}$. $\delta$ is nothing but the
defect of the hyperbolic triangle $\triangle ABC$ plotted in
Fig.\ref{fig}.
 And $\cos^2 \frac{\delta}{2}$ is proved as an upper bound of the Bures fidelity between two states.
 We also discussed the definitions of metrics based on the A-fidelity. And the numerical result sustains our expectation.

In Ref. \cite{ChenPRA2}, the authors have introduced an alternative
fidelity for a qunit holding the same  geometric observation as the
qubit system. Their precept is to define the fidelity of a qunit in
terms of the Bloch vectors in the formula of the qubit case. We can
insert the Bloch vectors into Eq. (\ref{GeoFa}), and obtain
\begin{eqnarray}\label{FaBl}
F_A(\rho_1,\rho_2)=\frac{[(1+\sqrt{1-|\mathbf{u}|^2})(1+\sqrt{1-|\mathbf{v}|^2})+\mathbf{u} \cdot \mathbf{v}]^2}{4(1+\sqrt{1-|\mathbf{u}|^2})(1+\sqrt{1-|\mathbf{v}|^2})}. \ \
\end{eqnarray}
An alternative A-fidelity for a $N$-state system can be defined as
\begin{eqnarray}\label{NFa}
&&\mathcal{F}_A(\rho,\sigma)= \nonumber \\
&&\frac{[(1+\sqrt{1-g(\rho, \rho)})(1+\sqrt{1-g(\sigma, \sigma)})+g(\rho, \sigma)]^2}{4(1+\sqrt{1-g(\rho, \rho)})(1+\sqrt{1-g(\sigma, \sigma)})}, \ \ \ \ \  \
\end{eqnarray}
where
\begin{eqnarray}
g(\rho,\sigma)=\frac{N \tr(\rho \sigma)-1}{N-1}.
\end{eqnarray}
Substituting Eq. (\ref{rhoN}) into Eq. (\ref{NFa}), it is easy to
prove that $\mathcal{F}_A(\rho,\sigma)$, which takes the form in Eq.
(\ref{FaBl}), can also be written in the form as Eq. (\ref{GeoFa}).
Therefore, it holds the same geometric picture as the fidelity of
the qubit system, and is a operable definition for a $N$-state
system.

\begin{acknowledgments}
This work is supported by the New teacher Foundation of Ministry of
Education of P.R.China (Grant No. 20070248087). J.L.C is supported
in part by NSF of China (Grant No. 10605013), and Program for New
Century Excellent Talents in University, and the Project-sponsored
by SRF for ROCS, SEM. F.L.Z thank Ning Ou-Yang for her help in
grammar.
\end{acknowledgments}

\bibliography{GeoFa}
\end{document}